\documentclass[pra,aps,showpacs,twocolumn,floatfix]{revtex4}
\usepackage{epsfig} \usepackage{graphics} \usepackage{bm} \usepackage{amssymb} \usepackage{color}
\begin{document}


\title{Fermionic stabilization and density-wave ground state of a polar condensate}
\author{O.~Dutta, R.~Kanamoto and P.~Meystre}
\affiliation{Department of Physics and College of Optical
Sciences, The University of Arizona, Tucson, AZ 85721, USA}
\date{\today}


\begin{abstract}
We examine the stability of a trapped dipolar condensate mixed
with a single-component fermion gas at $T=0$. Whereas pure dipolar
condensates with small $s$-wave interaction are unstable even for
small dipole-dipole interaction strength, we find that the
admixture of fermions can significantly stabilize them, depending
on the strength of the boson-fermion interaction. Within the
stable regime we find a region where a ground state is
characterized by a density wave along the soft trap direction.
\end{abstract}
 \maketitle


The recent demonstration of a condensate of chromium atoms
\cite{chbec} opens up the study of quantum-degenerate gases that
interact via long range, anisotropic magnetic dipole interactions.
In a parallel development, it can be expected that quantum
degenerate samples of heteronuclear polar molecules will soon be
available through the use of Feshbach resonances \cite{Kett,
Jin1}, photoassociation \cite{DeM, HetDeM1}, or a combination of
the two approaches, other possible routes including buffer-gas
cooling \cite{buff}, collisional beam cooling \cite{coll},
reactive scattering \cite{reac} and stark deceleration
\cite{stark} alone or in combination. When in their vibrational
ground state these molecules interact primarily via the electric
dipole interaction, and may therefore also form dipole-dominated
condensate.

As a result of the dipole-dipole interaction, a number of novel
phenomena have been predicted to occur in low-density
quantum-degenerate atomic and molecular systems, both in
conventional traps and in optical lattices. The existence of a
variety of quantum phases including a ``supersolid'' phase for
dipolar bosons in optical lattices have been predicted in
Ref.~\cite{oplat}. A supersolid phase was also predicted for
hardcore bosons in two-dimensional triangular
lattices~\cite{tri1,tri2}, in Kagome lattices~\cite{frus}, and in
an extended multi-band Bose-Hubbard Hamiltonian~\cite{exten},
although recent quantum Monte Carlo simulations \cite{monte} did
not find any supersolid phase for bosons in a Kagome lattice. The
authors of Ref.~\cite{bsfer} studied Bose-Fermi mixtures in
two-dimensional square lattices, and found a bosonic supersolid
transition induced by a modulation of the fermionic density
resulting from a nesting effect. However, the situation is much
different in the absence of a lattice structure: the only stable
state of dipolar condensates in a pancake-shaped trap has a
Gaussian-like density profile~\cite{bohn, rotmax, dpstab}, and
states with a periodic density modulation are always
unstable~\cite{coop}, although Ref.~\cite{odell} has shown that
modulating periodically the strength of a laser-induced
dipole-dipole interaction can result in stable density-modulated
condensates.

In this note we consider a mixture of dipolar bosons and
non-interacting single-component fermions confined to a
cigar-shaped trap. Such a mixture might be realized using isotopes
of chromium, or during the production of heteronuclear molecules
via either a Feshbach resonance or the photoassociation of two
different fermionic species of atoms. Using linear response theory
to determine the fermion-induced interaction between bosons we
find that it significantly stabilizes the dipolar bosonic
condensate. We then identify a region in parameter space where the
stable ground state displays a density-modulated structure along
the long axis of the trap.

We assume for simplicity that both the dipolar bosons and the
fermions are trapped in the transverse direction
by a tight harmonic potential of frequency $\omega_{\bot}$ and in the
longitudinal direction by a much softer harmonic potential of
frequency $\omega_z$. The dipoles are taken to be polarized by an
electric or magnetic field in a direction $y$ perpendicular to the
long axis of the trap. The dipole-dipole interaction between two
bosonic particles separated by a distance $r$ is given by $V_{\rm
dd}(r) = g_{\rm dd} \left ( 1 - 3 y^2/r^2 \right )/{r^3}$, where
$g_{\rm dd}$ is the dipole-dipole interaction strength. In the
mean-field approximation, the energy functional for the order
parameter $\phi(\bm{r})$ of the dipolar condensate is
\begin{eqnarray}\label{energy}
E\!\!&=&\!\!\!\int \phi^{*}(\bm{r}) H_0 \phi(\bm{r}) d^3r + \frac{Ng}{2}
\int | \phi(\bm{r}) |^4 d^3r \\
&+&\!\!\frac{N}{2} \! \int\!\!\!\!\int \! | \phi(\bm{r})  |^2
V_{\rm dd}(\bm{r} - \bm{r}') | \phi(\bm{r}') |^2 d^3r d^3r' \!+\!
E_{\rm ind}, \nonumber
\end{eqnarray}
where $H_0 = -\hbar^2\nabla^2/(2 m)  + m \omega^2_{\bot} \left
(x^2 + y^2 + \lambda^2 z^2 \right )/2$ is the sum of the kinetic
energy and the trapping potential, $m$ is the mass and $N$ the
total number of bosonic particles, and $\lambda=
\omega_z/\omega_{\bot}$. the second term in $E$ is the boson-boson
$s$-wave interaction of strength $g$, and the third term describes
the nonlocal dipole-dipole interaction between bosons. Note that
that term also contains a short-range contact contribution
~\cite{sc1, sc2, sc3}. The last term $E_{\rm ind}$ accounts for
the fermion-induced interaction $V_{\rm ind}(\bm{k})$ between
bosons. It is given by
\begin{equation}\label{fermi}
E_{\rm ind}=\frac{1}{2} \frac{g_{\rm bf} N}{(2\pi)^3} \int V_{\rm
ind}(\bm{k}) n(\bm{k}) n(-\bm{k}) d^3k,
\end{equation}
where $n(\bm{k})$ is the momentum-space bosonic density.

In deriving Eq.~(\ref{fermi}) we have assumed a contact
boson-fermion interaction of strength $g_{\rm bf}$, with a
corresponding interaction energy of the form $g_{\rm bf} \int
n_f(\bm{k}) n(-\bm{k})d^3k$, $n_f(\bm{k})$ being the fermion
density. The linear response of the fermions to a bosonic density
fluctuation $n(\bm{k})$ can be expressed as $n_f(\bm{k})= V_{\rm
ind}(\bm{k})n(\bm{k})$, and Eq.~(\ref{fermi}) is obtained by
substituting that expression back into to the boson-fermion
interaction energy. The explicit form of the induced potential is
$V_{\rm ind}(\bm{k})= g_{\rm bf}\chi_f(\bm{k})$, where $\chi_f$ is
the density response function~\cite{qub} related to the dynamical
structure factor $S(\bm{k},\omega)$, the probability of exciting
particle-hole pairs with momentum $\bm{k}$ out of the Fermi sea,
by $\chi_f(\bm{k})=-2 \int^{\infty}_0
d\omega'[S(\bm{k},\omega')/\omega'].$ For a non-interacting
single-component Fermi system we have
\begin{equation}\label{struc}
S(\bm{k},\omega)= \mathop{\sum_{p<k_f}^{\infty}}_{|\bm{p}+\bm{k}|>k_f}
\delta(\omega-\omega^0_{\bm{p} \bm{k}}).
\end{equation}
Here $p=|\bm{p}|$, $k_f$ is the Fermi momentum, and the excitation
energy is $\omega^{0}_{\bm{p} \bm{k}}=pk \cos \theta/m_f
+k^2/(2m_f)$, $\theta$ being the relative angle between $\bm{p}$
and $\bm{k}$, and $m_f$ the mass of the fermions. This expression
assumes that the fermions are locally free, so that there is a
Fermi sphere in momentum space.

Using the form of $S(\bm{k},\omega)$ from Ref.~\cite{qub} we find
\begin{equation}\label{indu1}
\chi_f(k) = -\frac{\nu k_f}{2k} \left [ \frac{k}{k_f}+\left (
1-\frac{k^2}{4k^2_f} \right )
\ln \left | \frac{2k_f+k}{2k_f-k} \right | \right]
\end{equation}
where $\nu$ is the three-dimensional density of states at the
Fermi surface. Here $|..|$ is used to denote absolute value.
Substituting Eq.~(\ref{indu1}) into the expression $V_{\rm
ind}(\bm{k})= g_{\rm bf}\chi_f(\bm{k})$ and carrying out a series
expansion to lowest order in the non-local terms gives
\begin{eqnarray}\label{indu}
V_{\rm ind}(k) \approx
\left\{
\begin{array}{lll}
\displaystyle{g_{\rm bf}\nu\left [ -1 + \frac{1}{3} \left ( \frac{k}{2 k_f} \right )^2 \right]},
&& k < 2k_f,\\
\displaystyle{\frac{g_{\rm bf} \nu}{12} \left ( \frac{2 k_f}{k} \right )^2},
&& k > 2k_f.
\end{array}
\right.
\end{eqnarray}
The negative contribution to this potential describes an effective
attractive interaction between bosons in the long wavelength
limit, while its positive part corresponds to a nonlocal repulsive
interaction.

With the form (\ref{energy}) of the bosonic energy functional and
the potential $V_{\rm ind}$ at hand, we now proceed to determine
the stability of the dipolar condensate, using a variational wave
function in the parameter space of the fermion-induced interaction
and dipolar strength. For this purpose, the condensate order
parameter $\phi(\bm{r})$ is assumed to factorize as $\phi(\bm{r})=
\phi_{\bot}(x, y) \phi_{\|}(z)$ with the normalization $ \int d^3
r| \phi(\bm{r})|^2=1$. The transverse wave function is
$\phi_{\bot}(x, y) = \exp [ - \left( x^2 + y^2 \right ) /(2 d^2)
]/\sqrt{\pi d^2}$, $d$ being a variational parameter, while the
longitudinal component is $\phi_{\|}(z)=\exp[-z^2/(2
d_z^2)]/\sqrt{\pi^{1/2}d_z}$, $d_z$ being likewise a variational
parameter.

\begin{figure}[ht]
\begin{center}
\epsfig{file=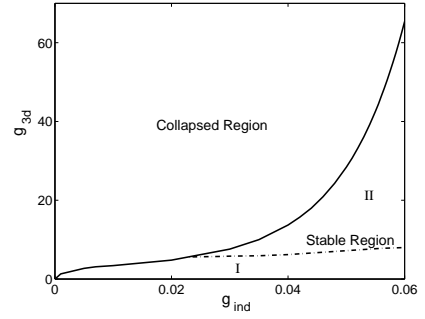,width=5.5cm} \caption{\label{fig1}
 Stability diagram of a dipolar condensate in ($g_{\rm 3d},g_{\rm ind}$) space
for $g_s=0.2$ and $\lambda=0.05$. In region I the stable ground
state has a Gaussian-like density distribution and in region II it
is characterized by a density modulation, i.e., $a_{\rm dw}>0$. }
\end{center}
\end{figure}

Substituting this Gaussian ansatz and its Fourier transform into
Eqs.~(\ref{energy}) and (\ref{fermi}),
the energy of the condensate becomes
\begin{eqnarray}\label{enfun}
\frac{2 m d^2_{\bot}}{\hbar^2} E_g \!\!\!&=&\!\!\!\left ( 1+\frac{1}{2}
\eta^2 \right ) \frac{d^2_{\bot}}{d^2} +
\left ( 1 + \frac{\lambda^2}{2 \eta^2} \right ) \frac{d^2}{d^2_{\bot}} \\
+ \ {g_{\rm 3d}}\frac{d^3_{\bot}}{d^3} \!\!\!\!\!\!&& \!\!\!\!\!\!
\left [ (g_s-1) \eta + \eta^3 \left \{ F(\eta) + g_{\rm ind}
\frac{d^2_{\bot}}{d^2} + \frac{g_{\rm ind} \tilde{k}^2_f}{\eta^2}
\right \}  \right ],\nonumber
\end{eqnarray}
where $\eta=d/d_z$, $\tilde{k}_f=k_f d_{\bot}$, $d_{\bot}=\sqrt{\hbar/m \omega_{\bot}}$
is the transverse oscillator length, and
\begin{equation}
F(x)=\frac{\tan^{-1}(\sqrt{x^2-1})}{(x^2-1)^{{3}/{2}}}-\frac{1}{x^2
(x^2-1)}.
\end{equation}
The first term in Eq.~(\ref{enfun}) denotes the short range part
of $V_{\rm dd}$, and the second term results from the difference
between the boson-boson $s$-wave interaction and the attractive
part of the interaction in $V_{\rm ind}(k)$. To calculate the
fermion-induced interaction energy $E_{\rm ind}$ we have assumed
that $2\tilde{k}_f>1$. In Eq.~(\ref{enfun}) we have introduced the
effective three-dimensional dipole-dipole interaction $g_{\rm
3d}=N g_{\rm dd} m/(\sqrt{2 \pi} \hbar^2 d_{\bot})$, the effective
contact potential $ g_s=1/3+(g-g_{\rm bf}^2\nu)/(2\pi g_{\rm
dd})$, and the induced interaction $g_{\rm ind}=g^2_{\rm bf}
\nu/(12 \pi g_{\rm dd}\tilde{k}^2_f)$.

The energy functional~(\ref{enfun}) is minimized with respect to
$d/d_{\bot}$ and $\eta = d/d_z$ for various combinations of the
parameters $g_{\rm 3d}$, $g_{\rm ind}$. The results of these
simulations are summarized in Fig.~\ref{fig1}. All numerical
results presented in the following are for $\lambda=0.05$,
$g_s=0.2$, and $\tilde{k}_f =2$ (a value typical of the situation
for a fermionic gas of $10^3$ atoms in a a transverse trap length
of $d_{\bot}\sim 1 \mu$m and $\lambda=0.05$). We found that the
condensate typically collapses for  $\eta \gtrsim 1$ or $d
\rightarrow 0$, while the stable regime is characterized by $ \eta
\lesssim 0.1 $ and $d \sim d_{\bot}$. Region I is characterized by
stable ground states with a Gaussian-like density profile, while
in Region II the ground state exhibits density modulations that
are further discussed later on.

In the absence of fermion-induced effective boson-boson
interaction $g_{\rm ind}=0$, the dipolar condensate is found to be
unstable to collapse even for very small dipolar strength $g_{\rm 3d}$.
For $g_{\rm ind}\neq 0$, in contrast, the condensate
becomes stable for a range of values of $g_{\rm 3d}$ that
increases with increasing $g_{\rm ind}$. This can be understood
from the form of the non-local part of the induced interaction,
Eq.~(\ref{indu}). Since the induced interaction $V_{\rm ind}$ is
repulsive for high momenta, see Fig.~3, the non-local repulsive
part balances the attractive non-local part of the dipole-dipole
interaction with increasing $g_{\rm ind}$, which results in
increased stability of the condensate. We confirmed that the
qualitative features of the stability diagram remain unchanged for
$0.05 \lesssim g_s \lesssim 0.3$. The system is also found to
become stable for higher dipolar strengths $g_{\rm 3d}$ with
increasing $g_s$.

We now restrict our considerations to the case of a cylindrical
trap, and discuss the appearance of a density-modulated ground
state in region II of the stability diagram. We proceed by first
demonstrating the existence of a roton minimum in the Bogoliubov
excitation spectrum of a homogeneous condensate along the trap
axis, and show that by increasing the effective dipolar strength
$g_{\rm 3d}$ this minimum can touch the zero-energy axis. The
roton-like spectrum follows from the attractive nature of dipolar
interaction for high momenta \cite{bohn, rotmax, rot1, fis} and
touching the zero energy indicates an instability toward a
nonuniform ground state \cite{rica}.

\begin{figure}[t]
\begin{center}
\epsfig{file=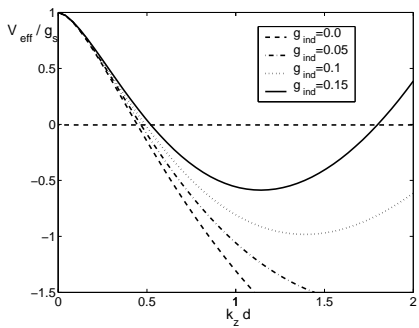,width=5.5cm} \caption{\label{fig2} Effective
one-dimensional potential, Eq.~(\ref{pot}), for several values of
the fermion-induced interaction $g_{\rm ind}$. Here we assumed
that $d \sim d_{\bot}$. The zero potential level is shown as a
horizontal dashed line.}
\end{center}
\end{figure}

Integrating Eqs.~(\ref{energy}) and (\ref{fermi}) in the
transverse direction yields the effective longitudinal potential
\begin{equation}\label{pot}
V_{\rm eff}(\tilde{k}_z) \!=\!
g_s \!-\! \frac{\tilde{k}^2_z}{2}
\exp\!\!\left(\frac{\tilde{k}^2_z}{2}\right)
E_1\!\!\left(\frac{\tilde{k}^2_z}{2}\right)
\!+\! {g_{\rm ind}}\frac{d^2_{\bot}}{d^2} \tilde{k}^2_z,
\end{equation}
where $\tilde{k}_z=k_z d$ is a scaled wave vector, $k_z$ being the
longitudinal momentum, and $E_1(x)=\int^{\infty}_x dx (e^{-x}/x)$
is the exponential integral. In obtaining Eq.~(\ref{pot}) we have
assumed that $\tilde{k}_z<\tilde{k}_f$ which justifies keeping
only the contribution to the induced potential (\ref{indu}) with
$k<2k_f$. Figure~\ref{fig2} shows that as $g_{\rm ind}$ is
increased the one-dimensional potential becomes less attractive
for higher momenta, and also that the non-local part of $V_{\rm
eff}(\tilde{k}_z)$ becomes repulsive again for sufficiently strong
$g_{\rm ind}$.

\begin{figure}[t]
\begin{center}
\epsfig{file=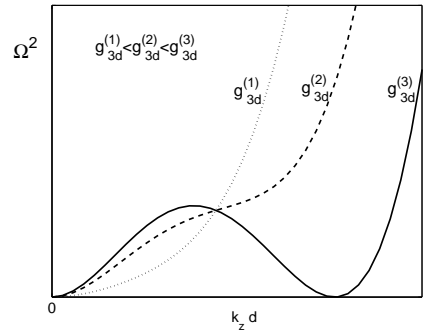,width=5.5cm} \caption{\label{fig3} Qualitative
diagram of the Bogoliubov spectrum for the effective longitudinal
potential from Eq.~(\ref{pot}) for several values of $g_{\rm 3d}$
and an arbitrary finite value of $g_{\rm ind}$. }
\end{center}
\end{figure}

Assuming a homogeneous condensate of length $L$ along the long
axis of the trap then yields the Bogoliubov spectrum
$$
\Omega^2(\tilde{k}_z) = \frac{\tilde{k}^2_z}{2} \left [ \frac{\tilde{k}^2_z}{2} +
\frac{2 g_{\rm 3d}}{L} V_{\rm eff}(\tilde{k}_z) \right ],
$$
with the appearance of a roton minimum for large enough effective
dipolar strength, see Fig.~\ref{fig3}.
By increasing $g_{\rm 3d}$ further,
the roton minima can touch the zero-energy axis.
This suggests that at this point the structure of the condensate
ground state can undergo a transition, with the appearance of a density
modulation along the long axis of the trap.

To verify this possibility we introduce a new longitudinal
variational wave function that includes a density-wave component
of amplitude $a_{\rm dw}$
\begin{equation}
\phi_{\|, {\rm dw}}(z) = \phi_{\|}(z)
\left [ a+a_{\rm dw} \cos\!\left(\frac{\tilde{k_0} z}{d}\right) \right ],
\end{equation}
with the normalization constraint $a^2+a^2_{\rm dw}/2=1$. The
excess energy of this density-modulated state, as compared to
a broad Gaussian order parameter, can be expressed as
\begin{eqnarray}\label{denfun}
\epsilon(\tilde{k}_0, a_{\rm dw})
\!\!\!&=&\!\!\!\frac{2  m d^2}{\hbar^2}(E_{\rm dw} - E_{g}) \\
=\frac{a^2_{\rm dw}}{2}\!\!\!\!\!\!&&\!\!\!\!\!\!\! \left
[\frac{\tilde{k_0}^2}{2} \!+\! \sqrt{2} g_{\rm
3d}\frac{d_{\bot}}{d_z} \!\left \{ 2 a^2 V_{\rm eff}(\tilde{k}_0)
\!+\! \frac{a^2_{\rm dw}}{8} V_{\rm eff}(2\tilde{k}_0)\! \right
\}\! \right ]. \nonumber
\end{eqnarray}

\begin{figure}[t]
\begin{center}
\epsfig{file=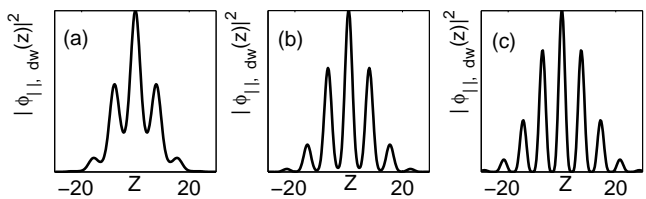,width=8.6cm} \caption{\label{fig4} Longitudinal
density profile of the condensate as a function of increasing
dipolar strength $g_{\rm 3d}$ for $g_{\rm ind}=0.058$ inside
region II of Fig.~1. The longitudinal displacement $z$ is scaled
to the transverse oscillator length $d_{\bot}$. (a) $g_{\rm
3d}=7.9$, where the system just enters region II, with excess
energy $\epsilon=-1.0\times 10^{-3}$, (b) $g_{\rm 3d}=12$,
$\epsilon=-3.5\times 10^{-2}$, and (c) $g_{\rm 3d}=20$,
$\epsilon=-2.0\times 10^{-1}$.}
\end{center}
\end{figure}

Limiting ourselves to the stable region of Fig.~1, we first choose
the values of $d$ and $ d_z$ that minimize the
energy~(\ref{enfun}), and then minimize $\epsilon(\tilde{k}_0,
a_{\rm dw})$ with respect to the variational parameters $k_0$ and
$ a_{\rm dw}$. Density-wave ground state is characterized by min
$\{ \epsilon \} < 0$ and $a_{\rm dw} \not=0$. The region of
density-wave ground state determined in this fashion corresponds
to region II of Fig.~\ref{fig1}. For $g_{\rm ind} \ge 0.023$, a
transition from a Gaussian-like state to a density-wave state is
observed when the dipole-dipole interaction $g_{\rm 3d}$ is
increased. This transition is smooth in the sense that $a_{\rm
dw}$ increases gradually from zero as we increase the dipolar
strength $g_{\rm 3d}$ beyond the critical value. Figure~\ref{fig4}
illustrates the gradual emergence of density waves for a fixed
induced strength $g_{\rm ind}=0.058$. We find that by gradually
increasing $g_{\rm 3d}$ to a modest value the density profile
acquires periodic zeros along the longitudinal direction.

One possible candidate to observe the predicted density wave
region is a mixture of bosonic$^{52}$Cr and fermionic $^{53}$Cr
atoms. Bosonic Chromium has a magnetic dipole moment of $6 \mu_B$
and a $s$-wave scattering length $\sim 100a_0$ \cite{ch, ch1}. The
trap considered in this paper has a transverse frequency
$\omega_{\bot} \sim 200$Hz and aspect ratio $\lambda=.05$. For
this geometry the density wave state in region II of
Fig.~\ref{fig1} can be reached for a boson-fermion scattering
length of $\sim 500a_0$, for bosonic and fermionic atoms numbers
of $10^4$ and $10^3$, respectively.

In summary, we have analyzed the stability of ultracold dipolar
bosons mixed with non-interacting fermions. A central result of
our analysis is that the fermions help stabilize the dipolar
condensate, a consequence of the fact that the non-local induced
interaction is repulsive in the limit of moderate wavelengths. In
the stable region we found a transition in the shape of the
condensate ground-state from a Gaussian-like profile to a
modulated density profile along the axis of the trap. Our analysis
is mean-field, implying that the system is assumed to be condensed
with phonon-like low energy excitations along the longitudinal
direction. The density modulation emerges as a result of the
additional breaking of the translational symmetry of the
condensate and can be called a superfluid density wave.

Future work will discuss the effect of the strength of the contact
interaction and of the Fermi momentum on the stability of the
condensate and the density wave, with possible extensions to
pancake geometries and to rotating systems. A number of exotic
states are likely to be found in that regime. An extension to
finite temperatures will allow us to investigate possible
classical phase transitions between superfluid and various kinds
of density-wave states \cite{denw}.

This work is supported in part by the US Office of Naval Research,
by the National Science Foundation, and by the US Army Research
Office.

\end{document}